\documentclass{article}

\usepackage{arxiv}
\usepackage{float}
\usepackage[T1]{fontenc}    
\usepackage{hyperref}       
\usepackage{url}            
\usepackage{graphicx}
\graphicspath{ {./images/} }
\usepackage{multirow}
\usepackage{xcolor}
\usepackage{array}
\usepackage{mathtools}
\usepackage{amsmath}

\usepackage{eqexpl}
\usepackage{empheq}
\usepackage{xcolor}
\definecolor{lightgreen}{HTML}{90EE90}
\eqexplSetDelim{=}

\newcolumntype{P}[1]{>{\centering\arraybackslash}p{#1}}
\definecolor{OliveGreen}{rgb}{0,0.6,0}
\definecolor{OliveBlue}{rgb}{0,0,0.6}
\definecolor{OliveRed}{rgb}{0.6,0,0}
\definecolor{AllOlive}{rgb}{0.6,0.6,0.6}
\definecolor{pink}{rgb}{0.8,0.6,0.2}
\definecolor{yellow}{rgb}{1,1,0}
\definecolor{purple}{rgb}{0.8,0.4,1}
\title{Differential Privacy for Credit Risk Model }

\usepackage[acronym]{glossaries}

\newacronym{dp}{DP}{Differential Privacy}
\newacronym{dpm}{DPM}{Differentially Private Model}
\newacronym{ndpm}{NDPM}{Non-Differentially Private Model}
\newacronym{pd}{PD}{Probability of Default}
\newacronym{ead}{EAD}{Exposure at Default}
\newacronym{lgd}{LGD}{Loss Given Default}

\date{} 					

\author {
\href{https://orcid.org/0000-0002-7895-6681}{\includegraphics{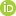}\hspace{1mm}Tabish Maniar}\\
Synechron Innovation Lab\\
\texttt{tabish.maniar1@synechron.com}\\
\And
\href{https://orcid.org/0000-0002-1538-9448}{\includegraphics{images/ORCIDiD_icon16x16.png}\hspace{1mm}Alekhya Akkinepally}\\
Synechron Innovation Lab\\
\texttt{alekhya.akkinepally@synechron.com}\\
\And
\href{https://orcid.org/0000-0002-9064-3362}{\includegraphics{images/ORCIDiD_icon16x16.png}\hspace{1mm}Anantha Sharma}\\
Synechron Innovation Lab\\
\texttt{anantha.sharma@synechron.com}\\
}

\hypersetup{
pdftitle={Differential Privacy for Credit Risk Model},
pdfsubject={cs.LG, stat.ML},
pdfauthor={Tabish Maniar, Alekhya Akkinepally,Anantha Sharma},
pdfkeywords={Differential Privacy, Credit Risk Model, Machine Learning},
}

\begin{document}
\maketitle
    
\begin{abstract}
The use of machine learning algorithms to model user behavior and drive business decisions has become increasingly commonplace, specifically providing intelligent suggestions/recommendations to automated decision making (like in loan applications etc.,). This has led to an increase in the use of customer’s personal data (such as age, SSN, and demographics, etc.,) to analyze customer behavior and predict their interests in a company’s products and/or services. Increased use of this customer personal data can   lead to better models but also to the potential of customer data being leaked, reverse-engineered, and mishandled.  

In this paper, we assess differential privacy as a solution to address these privacy problems by building privacy protections into the data engineering and model training stages of predictive model development. Our interest is pragmatic implementation in an operational environment, which necessitates a general-purpose differentially private modeling platform/framework, and we evaluate one such a tool from LeapYear as applied to the Credit Risk modeling domain. Credit Risk Model is a major modeling methodology in banking and finance where user data is analyzed to determine the total Expected Loss to the bank. We examine the application of differential privacy on the credit risk model and evaluate the performance of a Differentially Private Model (DPM) with a Non-Differentially Private Model (NDPM). 

Credit Risk Model is a major modeling methodology in banking and finance where user's data is analyzed to determine the total Expected Loss to the bank. In this paper, we explore the application of differential privacy on the credit risk model and evaluate the performance of a \gls{ndpm} with a \gls{dpm}.
\end{abstract}

\providecommand{\keywords}[1]{\textbf{\textit{}} #1}
\keywords{Differential Privacy, Credit Risk Model, Machine Learning}

\section{Introduction}

In the past few years, the use of machine learning algorithms to model user behavior has been prevalent. With this, the personal data of the user such as Location, Age, SSN have been met with several attacks both from internal as well external attack vectors. A significant internal vulnerability is the risk that the data scientist or others with access to personal data may accidentally or maliciously leak confidential information, and so there is a need to prevent this from happening while enabling legitimate development and workflows on the underlying data. And it is imperative that the trained model does not inadvertently expose information about the users in the training data, a well-known risk \cite{zhanglong}. 

This has led to research and applications in the area of Differential Privacy (DP). In this paper, we assess the application of \gls{dp} for Credit Risk Modeling, which deals with understanding and predicting risks associated with loan disbursement to one or more parties with multiple areas for risk modeling. Concretely, we perform  exposure modeling including the probability of default—i.e. the total amount of (potential) exposure at the time of default, how much the loan is expected to be worth at the time of default, and the overall loss if there is a default.

This paper deals with exposure modeling including the probability of default. ie., the total amount of (potential) exposure at the time of default, how much the loan is expected to be worth at the time of default, and the overall loss if there is a default.

The purpose of this study was to address two main research questions about using \gls{dp} to develop credit risk models:

\begin{itemize}
    \item Can privacy-preserving models perform as well as non-private models in terms of key outcomes such as loss estimation and variance? 
    \item What are the operational considerations of privacy-preserving models, such as computational overhead and implementation of trained models? 
\end{itemize}

\section{Brief look at Differential Privacy}
\gls{dp} Differential Privacy provides a mathematical basis for privacy protection, with provable limitations on the amount of information that can be extracted about a particular record in a database, such as a customer or transaction, regardless of what other information or computational resources are available to a would-be adversary \cite{Cynthia}. The mathematically provable privacy protection of \gls{dp} stands in contrast to approaches such as anonymization and masking, which are highly vulnerable to re-identification \cite{unique}. \gls{dp} can provide protection for two vulnerabilities in the context of predictive model development: a model developer who needs to leverage training data, including private information, for exploratory analysis and modeling can do so without direct access to the source data and with mathematical protection against reverse engineering of protected information; and the \gls{dpm} protects information leakage from individual records in the training data. 
Differential Privacy has been applied to multiple domains such as on census data \cite{ferdinando} where they have used Rappor as their differential privacy framework \cite{rappor}. US Census Bureau adopted differential privacy to protect the End-to-End Census Test \cite{uscensus}. It has been applied to obfuscate the images from dataset \cite{imageobfuscation} and recently the ImageNet has blurred the images in the dataset with minimal impact on the accuracy. \gls{dp} has applications in the space of Health where analysts have to analyze private health records \cite{healthdata} \gls{dp} has been applied to create Recommender Systems which can handle private datasets to protect the user’s history \cite{recommmender}. Differential Privacy can also apply to Location Datasets such as the path of taxis and users’ location history which is private information to the user and can likely be misused \cite{dpapplications}. 

\section{Credit risk models }
Credit risk measures the risk of a loss that may occur if a borrower does not make the required payments on the loans or fails to meet his or her debt repayment obligations. Credit risk modeling is a technique used by lenders to estimate the level of risk associated with each borrower. The risk for the lender is of several kinds, ranging from disruption to cash flows and increased collection costs to loss of interest and principal. That’s why it’s important to be able to anticipate credit risk as accurately as possible. Credit risk on a variety of individual characteristics (factors), and the nature of dependency could be quite complex. Lenders employ sophisticated models which excel at capturing this dependency. 

There are several major factors to consider while determining credit risk. From the financial health of the borrower and the consequences of a default for both the borrower and the creditor to a variety of macroeconomic considerations.

The credit risk of a borrower is typically quantified via three major components, which are modeled separately:
\begin{itemize}
\item\textbf{\gls{pd}:} This model gives if the borrower will default or not. It is the likelihood that the borrower will not be able to make scheduled repayments of the loan. When \gls{pd} is higher the lender will charge a higher rate. It is a classification model which would determine if the borrower will default or not within a specified time frame given KYC information, historical repayment data (of similar customers), and present and future economic outlook.
\item\textbf{\gls{ead}:} \gls{ead} is the maximum amount that the bank is exposed to, at the moment the borrower defaulted on the loan obligation. It is often modeled focusing on a credit conversion factor, which is defined as a ratio to the original loan amount, so that:
\begin{equation} \label{eq:1}
 \boxed{exposure\_at\_default = total\_funded\_amount * credit\_conversion\_factor}
\end{equation}
\begin{equation} \label{eq:2}
 \boxed{credit\_conversion\_factor=\frac{total\_funded\_amount* total\_recovered\_principal}{total\_funded\_amount}}
 \end{equation}
where:\\
\textbf{total\_funded\_amount} is the initial amount that was lend to the borrower\\
\textbf{credit\_conversion\_factor} is the proportion of the original funded amount that the borrower defaulted on. The borrower may have already payed back some part of the amount that has been initially borrowed.\\
\textbf{total\_recovered\_principal} is the amount that has already been payed back by the borrower before the  borrower has defaulted.\\
\item\textbf{\gls{lgd}:} \gls{lgd} is the amount which financial institute loses when a borrower defaults on a loan after recovery.It is the loss that can not be recovered after the default.  

\begin{equation} \label{eq:3}
 \boxed{loss\_given\_default=1-recovery\_rate}
\end{equation}
where:\\
\textbf{recovery\_rate} is the loss that can be recovered.\\

Using the output from the models we calculate the Total Expected Loss by using the formula. 

\begin{equation} \label{eq:4}
 \boxed{total\_expected\_loss = \sum (probability\_of\_ default *  exposure\_at\_default * loss\_given\_ default)}
\end{equation}
where:\\
\textbf{total\_expected\_loss} is the total amount that the bank is expected to loose on the all the borrowers.
\end{itemize}

\section{Methodology}
\subsection{Data}
To build the Credit Risk Model we used Lending Club dataset which is publicly available for the years 2016 and 2017. The dataset has 39K records and multiple features about the lender such as age, loan amount, the status of the loan, total recovered principal, recoveries, address, loan status \cite{lendingclub}. We are using this data to determine loss to the bank over a period of time when a customer defaults. We have split the data into train and test datasets with an 80:20 ratio.

\subsection{DP Software Selection}
A key objective of this study was to evaluate an operationally realistic end-to-end \gls{dp} workflow for developing credit risk models, to address privacy vulnerabilities for both the model development process (i.e., to protect data from the model developer) as well as implementation of the trained \gls{dpm} (i.e., to protect training data from end-users of the model). Furthermore, given that financial institutions utilize a broad range of data platforms and services, from on-premise servers to various cloud providers, it was desirable to evaluate software that was not limited to a particular proprietary computing environment. Other selection criteria included: support for \gls{dp} modeling algorithms relevant to credit risk modeling; computational scalability for rapid iteration; and ease of use, i.e., similarity to common non-DP analytics tools.

Though \gls{dp} is a very active area of research, today there are relatively few organizations and open source efforts that offer \gls{dp} software for practical applications. Some provide \gls{dp} algorithms for certain types of calculations (e.g., Google \gls{dp} \cite{google_dp}, IBM DiffPrivLib \cite{ibm_dp}) or tools focused on specific applications such as statistical releases (e.g., OpenDP \cite{openDP}), and yet others address different privacy scenarios than financial modeling, such as local differential privacy (e.g., Apple \cite{apple_dp}). 

The \gls{dp} analytics platform from LeapYear Technologies \cite{leapyear} is intended to support end-to-end private workflows like our use case. All interactions with the source data are differentially private, meaning the analyst performs data engineering, exploratory data analysis, feature generation, and model training interactively but without direct access to the data. Furthermore, privacy exposure is quantified and tracked for all interactions and actively managed to prevent information leakage across multiple uses of the same data. LeapYear thus meets the operational objectives to apply \gls{dp} to the entire workflow. LeapYear uses Apache Spark for open-source scalable computation, with a Python API that is similar to non-DP open-source analytics tools (e.g., PySpark). And it operates on Linux servers that are supported by a broad range of cloud and on-premise data environments.

\subsection{Pre-processing}
The source data was prepared for modeling with a similar process for \gls{dpm} and \gls{ndpm} modeling. These steps included:
\begin{itemize}
    \item Binning of states, home ownership variables, and purpose of the loan.
    \item Removal of highly correlated and unwanted columns from the dataset such as loan amount, zip code, and member id.
    \item Treatment of null values by imputing the median of the data attributes; and one-hot encoding of categorical variables to ensure compatibility across algorithms, some of which automatically handle categorical variables while others do not.
\end{itemize}

For \gls{dpm} development preprocessing was performed with the LeapYear platform, emulating the scenario where an analyst accomplishes this preprocessing without direct access to the source data (from a database or a data lake). Notably, such transformations are deterministic and exact in LeapYear because \gls{dp} randomization is applied at the point of computing answers to questions that analyst requests to see, not to the underlying data manipulations.

\subsection{Algorithms}
\begin{itemize}
\item\textbf{\gls{pd}:} The dependent variable for modeling \gls{pd} would be loan status which would determine if the borrower is good or bad. We have used multiple models to calculate \gls{pd} such as Logistic Regression and Gradient Boosting Trees. Comparatively Gradient Boosting Trees performed better.
\item\textbf{\gls{ead}:} The dependent variable for modelling \gls{ead} would be credit conversion factor(CCF) defined above (see \ref{eq:2}). To predict CCF we have used the Random Forest regression Model.

\begin{equation} \label{eq:5}
 \boxed{Predicted EAD = Total Funded Amount * predicted CCF}
\end{equation}

\item\textbf{\gls{lgd}:} The dependent variable for modeling \gls{lgd} is the recovery Rate. \gls{lgd} is simply (1 - Recovery Rate). Usually, \gls{lgd} is calculated using beta regression, which we have implemented by creating two models. A logistic regression classifier to model whether the recovery rate will be non zero and a linear regression model to predict the recovery rate when the recovery rate is non zero. For \gls{lgd}, we have used gradient boosted trees to model if the recovery rate is non zero and to predict the recovery rate we have used Random Forest Model.
\end{itemize}

\section{Result}
\subsection{Model Performance}
To see whether the \gls{dpm} would generate comparable results to traditional modeling approaches, we have implemented a process to a) train models on a sub-sample of the data and b) generate model predictions on the same training data sample. We have then executed this process several times on different data samples and compared the actual and model predicted loss across these data samples: 

\begin{table}[!htbp]
\centering
\resizebox{\textwidth}{!}{%
\begin{tabular}{|c|c|c|c|c|c|c|}
\hline
\multirow{2}{*}{Run} &
  \multicolumn{2}{c|}{Total Actual Loss} &
  \multicolumn{2}{c|}{Total Predicted Loss} &
  \multicolumn{2}{c|}{Relative Difference} \\ \cline{2-7} 
 &
  Exact Total &
  Differentially Private Total &
  Traditional Approach &
  LeapYear Differentially Private Model &
  Traditional Approach &
  LeapYear Differentially Private Model \\ \hline
1 & \$8,428,504 & \$8,405,516 & \$7,558,465 & \$7,787,930 & 11.510 & 7.930  \\ \hline
2 & \$7,848,538 & \$7,881,671 & \$6,851,392 & \$7,603,560 & 14.553 & 3.657  \\ \hline
3 & \$8,386,365 & \$8,407,715 & \$7,273,870 & \$6,812,098 & 15.294 & 23.423 \\ \hline
4 & \$8,282,985 & \$8,257,595 & \$7,045,931 & \$6,428,003 & 17.556 & 28.462 \\ \hline
5 & \$8,379,437 & \$8,382,792 & \$7,507,658 & \$6,884,052 & 11.611 & 21.771 \\ \hline
6 & \$8,232,474 & \$8,216,831 & \$6,563,581 & \$6,717,644 & 25.426 & 22.317 \\ \hline
7 & \$8,714,153 & \$8,684,676 & \$7,342,829 & \$6,998,088 & 18.675 & 24.100 \\ \hline
8 & \$8,285,473 & \$8,305,918 & \$6,730,521 & \$6,734,247 & 23.102 & 23.338 \\ \hline
\end{tabular}%
}
\caption{Individual Run Results}
\label{tab:results}
\end{table}

\begin{table}[!ht]
    \centering
    \begin{tabular}{|p{9em}|p{7em}|p{8em}|p{8em}|}
        \hline
        Model Version & Avg Actual Loss &  Avg Predicted Loss & Avg Relative Difference from Actual\\
        \hline
        Model trained in PySpark with full access to the data & \$8,319,741 & \$7,109,281 & 17.21 \\
        \hline
        Model developed via LeapYear with only differentially private compute access & \$8,317,839 & \$6,995,703 & 19.37
        \\ 
        \hline
    \end{tabular}
    \caption{Average Results}
    \label{tab:Average Results}
\end{table}
\newpage

As we can see in Table \ref{tab:Average Results} above, switching from a traditional modeling approach to a differentially private one - led to less than 5\% change in the aggregate expected loss prediction. This effect is well in line with the effect of choosing a different sub-sample for training models, as shown in Table \ref{tab:results} above and illustrated in \ref{fig:variance} below.
 
\begin{figure}[ht!]
    \centering
    \includegraphics[scale=0.5]{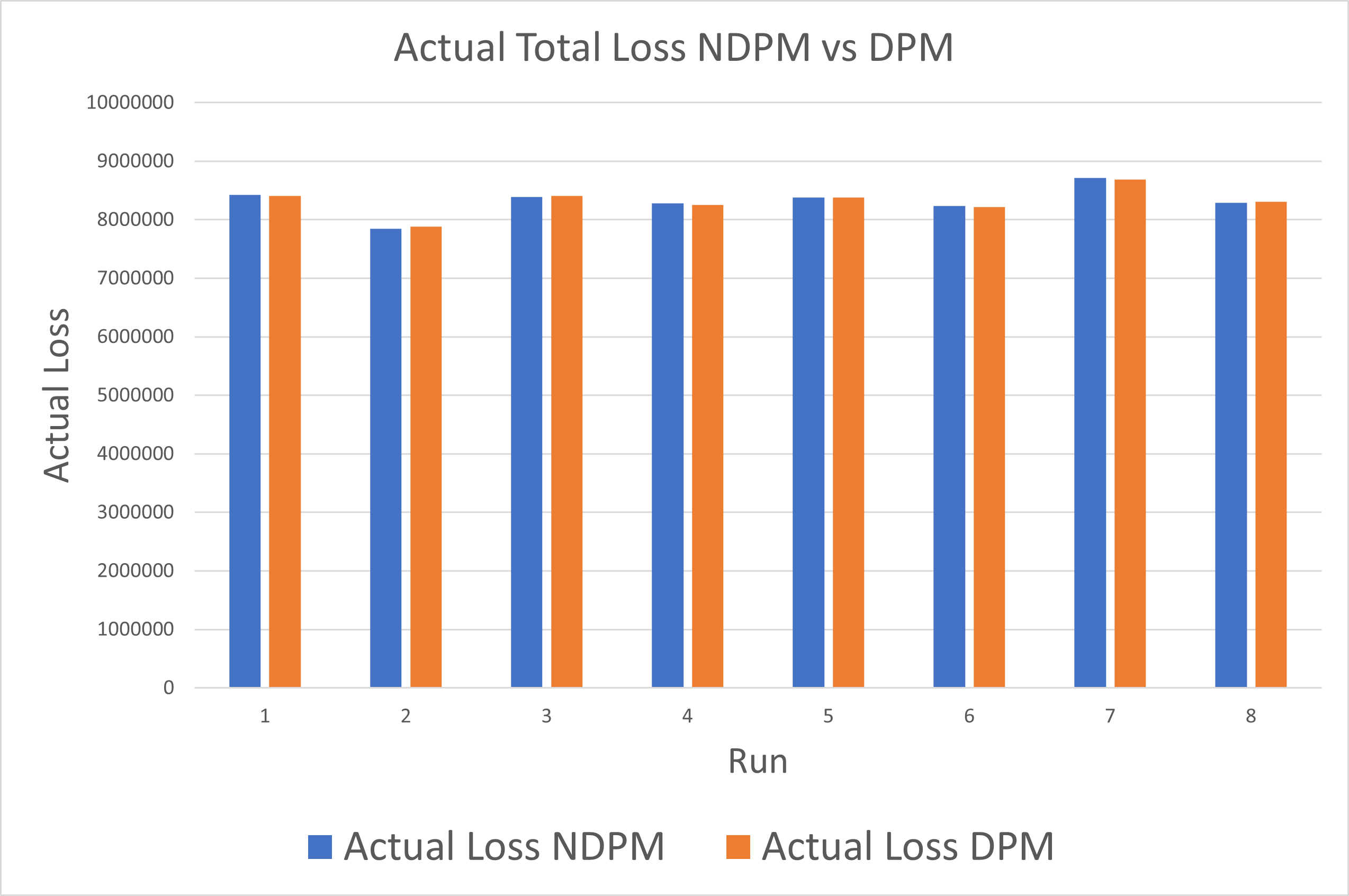}
    \caption{Actual Loss for \gls{dpm} and \gls{ndpm}}
    \label{fig:actual loss}
\end{figure}

\begin{figure}[ht!]
    \centering
    \includegraphics[scale=0.5]{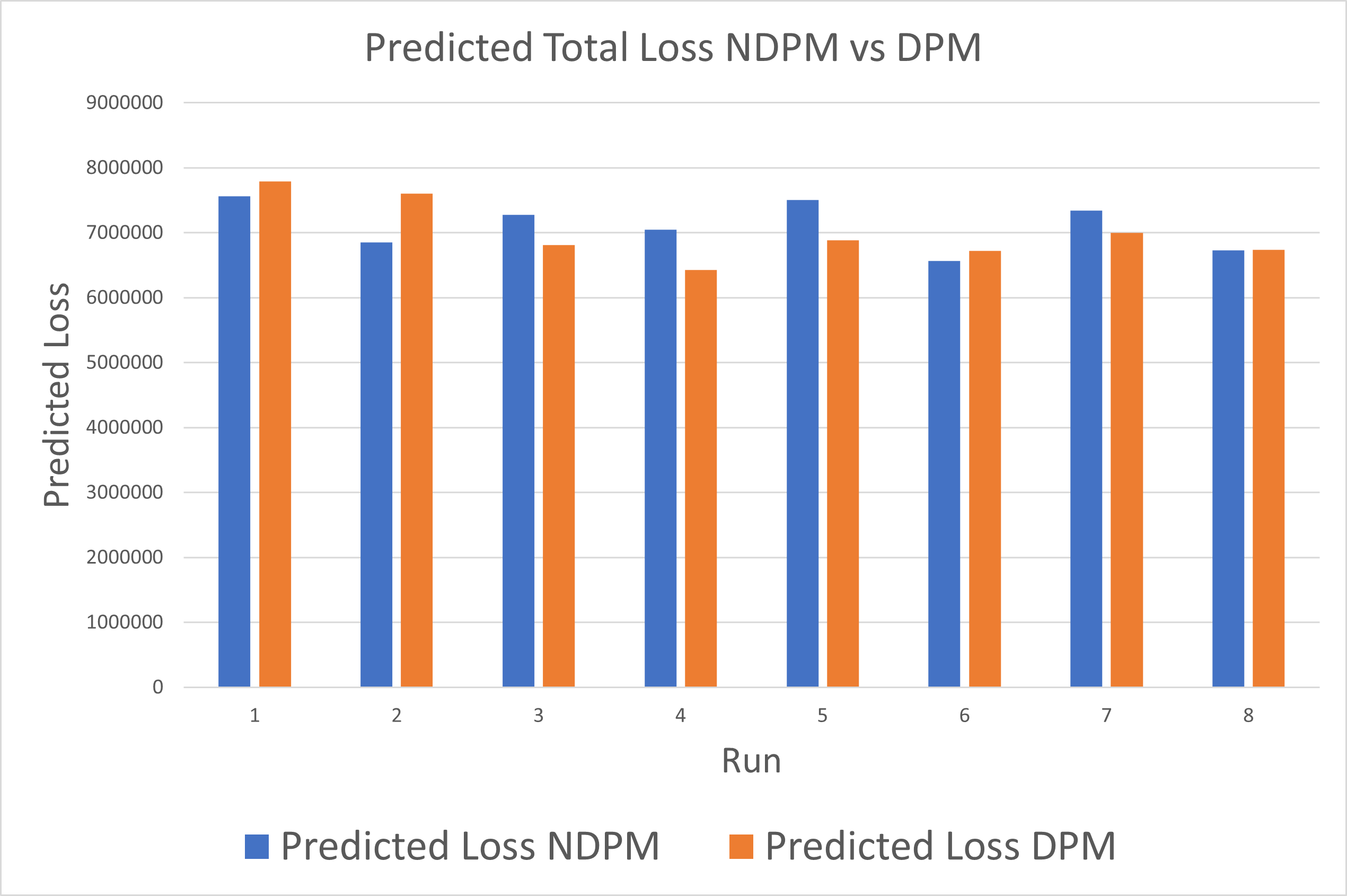}
    \caption{Predicted Loss for \gls{dpm} and \gls{ndpm}}
    \label{fig:Predicted loss}
\end{figure}
Figure \ref{fig:Predicted loss}. explains the different runs on the dataset with different sample of the dataset as training and testing. It compares the performance of each run of the \gls{dpm} and \gls{ndpm} model on the dataset.

\begin{figure}[ht!]
    \centering
    \includegraphics[scale=0.5]{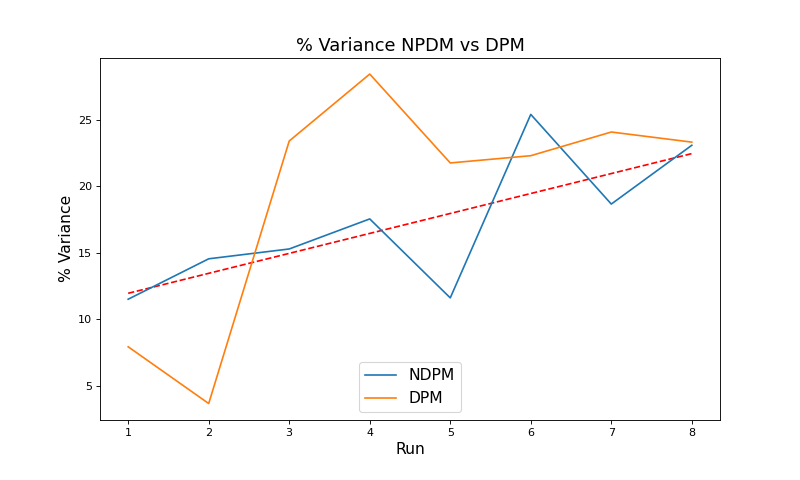}
    \caption{Variance \gls{dpm} vs \gls{ndpm}}
    \label{fig:variance}
\end{figure}

Figure \ref{fig:variance}. shows the variance from the actual value and predicted value in each run for the \gls{dpm} and \gls{ndpm}.
\raggedbottom

\subsection{Operational Considerations}

The processing time required to train the model was somewhat slower for the \gls{dpm}, 5.4 minutes, compared to the \gls{ndpm} model, 3.1 minutes. Prediction time was essentially the same between the two models. Pre-processing time, such as feature generation, was quite different between the private and non-private methods in this experiment, taking 13.5 minutes for the \gls{dpm} vs. 0.3 minutes for the \gls{ndpm}. LeapYear advised that feature calculations can be optimized to reduce or eliminate this difference, but this was not explored further in this study, and pre-processing time is generally less important than training and prediction time for operational implementation. 
It is possible to productionalize \gls{dpm} developed with Leapyear by first mapping the trained model to one of the common formats such as sklearn, pmml, and pfa. This process was demonstrated for linear regression and logistic regression in this study. More complex tree-based models can be similarly converted from LeapYear’s standard format to other formats in principle, but this was not demonstrated during this study. Training can be done in a Leapyear environment but the deployment of Leapyear models can be done by persisting the model to the storage disk and then loading the model in a non-LeapYear environment. 

Another desirable advantage of the \gls{dpm} is to keep the data hidden from the analyst during the model development process, making the sensitive data of the individual customers more secure. Note that this requires access controls in addition to \gls{dp}, as with the LeapYear platform. 

In addition to generating model predictions for individual records, these LeapYear model predictions for individuals could be explained and rationalized using similar tools as traditional open-source models - e.g. shap, see link: https://shap.readthedocs.io/en/latest/index.html. This workflow can be enabled by mapping models trained with LeapYear - to more traditionally supported formats such as scikit-learn.

\section{Conclusion}
In the context of the research question “Can privacy-preserving models perform as well as non-private models?”, it was observed that \gls{dpm} perform comparably to their \gls{ndpm} counterparts while providing added peace of mind for customers and banks on data privacy protection. Operationally, computational overhead for \gls{dpm} was negligible for model predictions, of comparable magnitude for model training, and noticeably larger for data pre-processing, though the operational impact can be mitigated or eliminated via automation and optimization of pre-processing. Production deployment of trained \gls{dpm} can be accomplished via conversion to standard model formats. Most importantly, \gls{dpm} can be developed without direct access to source data via a combination of \gls{dp} and access controls, as with the LeapYear platform. Additionally, the ability to add noise in a controlled manner specific to a given use case will allow banks to avoid data duplications in their data lake.

\raggedbottom
\bibliographystyle{unsrt}
\bibliography{bibliography.bib}
\end{document}